\documentclass[twocolumn,a4paper,showpacs,prl,aps]{revtex4}

\usepackage{color}
\usepackage{amsmath}
\usepackage{graphicx}
\usepackage{dcolumn}
\usepackage{bm}
\usepackage{comment}
\usepackage[caption=false]{subfig}
\usepackage{bbold}
\usepackage{bbm}

  \newcommand{\expval}[1]{\left< #1 \right>}
  
  \newcommand{\ii}{{\mathrm i}}

\newcommand{\ket}[1]{| #1 \rangle}
\newcommand{\bra}[1]{\langle #1 |}

\newcommand{\beq}{\begin{eqnarray}}
\newcommand{\eeq}{\end{eqnarray}}

\begin{document}

\title{Driven open quantum systems and Floquet stroboscopic dynamics} 

\author{S. Restrepo$^1$}
\author{J. Cerrillo$^1$}
\author{V. M. Bastidas$^2$}
\author{D. G. Angelakis$^{2,3}$}
\author{T. Brandes$^1$}
\affiliation{$^1$
Institut f\"ur Theoretische Physik, Technische Universit\"at Berlin, Hardenbergstr. 36, 10623 Berlin, Germany}%
\affiliation{$^2$
Centre for Quantum Technologies, National University of Singapore, 3 Science Drive 2, Singapore 117543}
\affiliation{$^3$School of Electronic and Computer Engineering, Technical University of Crete, Chania, Crete, Greece, 73100}

\date{\today}

\begin{abstract}

We provide an analytic solution to the problem of system-bath dynamics under the effect of high-frequency driving that has applications in a large class of settings, such as driven-dissipative many-body systems. Our method relies on discrete symmetries of the system-bath Hamiltonian and
provides the time evolution operator of the full system, including bath degrees of freedom, without weak-coupling or Markovian assumptions. An interpretation of the solution in terms of the stroboscopic evolution of a family of observables under the influence of an effective static Hamiltonian is proposed, which constitutes a flexible simulation procedure of non-trivial Hamiltonians. We instantiate the result with the study of the spin-boson model with time-dependent tunneling amplitude. We analyze the class of Hamiltonians that may be stroboscopically accessed for this example and illustrate the dynamics of system and bath degrees of freedom.   
\end{abstract}


\maketitle

\textit{Introduction:-}
An external driving breaks continuous translational invariance in time, which is associated with lack of energy conservation. In a seminal paper~\cite{Shirley1965}, Shirley showed that, in the case of a periodically-driven quantum system, the discrete translational symmetry in time can be exploited to derive a theory analogous to the Bloch theorem in condensed matter.
Despite the lack of energy conservation, one can still introduce a stroboscopic conserved quantity referred to as the quasienergy---similar to the quasimomentum in condensed matter. 
In Floquet theory, extension of the original Hilbert space of square integrable functions with one of periodic functions in time~\cite{Sambe1973} facilitates the identification of additional symmetries of the system that emerge as a consequence of the driving.

Driving-induced symmetries are a powerful tool in the theory of driven quantum systems~\cite{Grifoni1998,Bukov2015,Eckardt2015,Hanggi1991,Hanggi1992,Shevchenko20101,Bastidas2014,Bastidas2012}. Among various applications, it becomes possible to define a generalized parity symmetry in the extended Hilbert space of a driven qubit such that quantum degeneracies are exploited to suppress tunneling in a coherent manner~\cite{Hanggi1991,Hanggi1992}. This has important consequences in the context of driven open quantum systems~\cite{Grifoni1998,Hone2009, HanggiFluctuation, Weiss2016,Hong2015,Kohler1997,Szczygielski2013,Szczygielski2014,Hausinger2010,Thorwart2009,Grifoni2004}, where a nontrivial interplay of time dependent fields and dissipation takes place.

Up until now, the study of periodically driven open quantum systems has relied heavily on the Floquet-Markov \cite{Kohler1997,Grifoni1998,Szczygielski2013,Szczygielski2014,Hausinger2010} approach. This approach consists on deriving a weak coupling Born-Markov master equation~\citep{Breuer2007,Weiss2008} in the Floquet basis of the driven system. Conditions under which periodically driven open quantum systems thermalize have also been studied recently \cite{Shirai2015,Liu2015}. Besides this approach, there is an obvious interest in reaching beyond the weak coupling regime and the Markovian approximation.

In this letter, we provide an analytic solution of a driven open quantum system based on a perturbative expansion up to first order with respect to the period of the driving. The solution is valid in strong-coupling and non-Markovian scenarios~\cite{Cerrillo2014,Brandes2005,Brandes2004} and suits a large class of systems provided that the general system-bath Hamiltonian has certain driving-induced discrete symmetries. In particular, this may also be used for the study of driven-dissipative many-body systems.

With the analytical solution at hand, it is possible to interpret the evolution in terms of a stroboscopic sampling of a class of observables under the effect of a static, effective Hamiltonian. For open quantum systems, this may involve both system and bath degrees of freedom. Due to growing interest in quantum simulation of open systems by means of driven control \cite{Schindler2013, Mintert2015b}, this insight opens up an exciting new path of exploration. In this letter we first present the solution, specify the conditions required for its application and propose a way to simulate static Hamiltonians of open quantum systems, based on the stroboscopic description of driven systems. We then apply our results to the spin-boson model, which is a paradigmatic model in the theory of open quantum systems~\cite{Weiss2008,Breuer2007,Thorwart2007,Grifoni2004}.

\textit{Floquet theorem and high frequency expansions:-} 
The suitability of Floquet theorem for the study of periodically driven systems has been extensively established~\cite{Shirley1965,Grifoni1998,Bukov2015,Eckardt2015}. In its most general form, it states that the evolution operator associated to a time-dependent periodic Hamiltonian $\hat{H}(t+T)=\hat{H}(t)$ can be decomposed as 
\begin{align}\label{F_theo}
	\hat{U}(t,t_1) =e^{-\ii \hat{K}_{t_0}^{\text{F}}(t)}e^{-i \hat{H}_{t_0}^{\text{F}}(t-t_1)}e^{i \hat{K}_{t_0}^{\text{F}}(t_1)}
	,
\end{align}
where $\hat{K}_{t_0}^{\text{F}}(t)$ is the {\em stroboscopic kick operator} and $\hat{H}_{t_0}^{\text{F}}$ is the {\em Floquet Hamiltonian}. The parametric time dependence $t_1\leq t_0\leq t_1+T$ is associated to the start of the stroboscopic evolution. The stroboscopic kick operator inherits the periodicity of the Hamiltonian  $\hat{H}(t)$, such that $\hat{U}(t_0+nT,t_0) = e^{-i \hat{H}_{t_0}^{\text{F}}nT}$, with $n$ an integer and $\hat{K}^F_{t_0}(t_0)=\hat{K}^F_{t_0}(t_0+nT)=0$.

With the exception of some simple systems, it is impossible to find a closed form for the Floquet Hamiltonian and the stroboscopic kick operator. Nevertheless one may resort to high frequency expansions (HFEs)~\cite{Goldman2014,Bukov2015,Eckardt2015,Rahav2003,Itin2015,Mikami2015,Mananga2011,Kuwahara201696,Lopez2013} such as the well known Floquet-Magnus expansion \cite{Mananga2011, Kuwahara201696, Lopez2013}. The HFE is defined as a power series in  $1/\omega_\text{L}$, where $\omega_\text{L}$ is the frequency of the driving. This makes $\omega_\text{L}$ the dominant energy scale of the system, since it has to remain larger than any energy scale of the undriven model to support a suitable truncation of the HFE~\cite{Bukov2015,Eckardt2015}.
It is possible to define a unitary transformation that completely removes the dependence of $t_0$ from the HFE~\cite{Bukov2015,Eckardt2015}. This defines the {\em kick operator}  $\hat{K}(t)$ and the {\em effective Hamiltonian} $\hat{H}^{\text{F}}$ so that Eq.~\eqref{F_theo} may be rewritten as
\begin{align}\label{Effe_Evol}
	\hat{U}(t,t_1) =e^{-i \hat{K}(t)} e^{-i \hat{H}^{\text{F}} (t-t_1)} e^{i \hat{K}(t_1)}
	.
\end{align}
Note that, unlike the Floquet Hamiltonian, the effective Hamiltonian  $\hat{H}^{\text{F}}$ does not necessarily generate the stroboscopic evolution of the system, since $\hat{K}(t_0)$ may not vanish.  Expansions up to first order for both forms [\eqref{F_theo} and \eqref{Effe_Evol}] can be found in the supplemental material~\cite{SupplementalInfo}. The {\em effective} or {\em Van Vleck expansion}~\cite{Bukov2015,Eckardt2015} based on \eqref{Effe_Evol} takes an especially simple form which may be exploited in the analytical derivation of the evolution operator of a large class of systems.

\textit{Analytical solution:-} We consider a system-bath Hamiltonian of the form
\begin{align}
      \label{drivendephasingGeneral}
      	\hat{H}(t)= \hat{H}_{\text{S}}(t)+ \hat{H}_{\text{B}}
      	+  \hat{S} \, \hat{X} ,      
\end{align}
where $\hat{H}_{\text{S}}(t)=\omega_0\hat{S} +A\cos(\omega_{\text{L}}t)\hat{V}$ is the Hamiltonian of the system and $\hat{S}$ and $\hat{V}$ are time independent operators. Correspondingly, $A$ is the amplitude and $\omega_{\text{L}}$ is the frequency of the external driving with period $T=2\pi/\omega_{\text{L}}$. The operator
$\hat{H}_{\text{B}}=\sum_{k=1}^{N}  \omega_k \hat{a}_k^{\dagger} \hat{a}_k$  is the Hamiltonian of the bath with $N$ modes, $\hat{X}=\sum_{k=1}^{N} g_k \left(\hat{a}_k^{\dagger} +\hat{a}_k \right)$, and $\hat{a}^{\dagger}_k$, $\hat{a}_k$  creation and annihilation operators of the bath.
The case where $A=0$ can be solved analytically since it is a dephasing-type model~\cite{Weiss2008,Breuer2007}, where the populations of the system remain stationary while the coherences decay. The external driving breaks the integrability of the model, but it also generates new symmetries in the Sambe space~\cite{Sambe1973}. For example, the Hamiltonian of Eq.~\eqref{drivendephasingGeneral} is invariant under $t\mapsto -t$, and
the combined action of the transformations $t\mapsto t+T/2$ and $\hat{V}\mapsto -\hat{V}$.

We now briefly explain how to obtain an analytical form of the reduced density matrix of the system from Hamiltonian \eqref{drivendephasingGeneral}. For further detail the reader is referred to the supplemental material ~\cite{SupplementalInfo}. We begin by going to a rotating frame defined by operator
$
\hat{\mathcal{U}}(t)= e^{  -i \frac{A}{\,\omega_{\text{L}}} \sin(\omega_{\text{L}} t)\, \hat{V}}.
$
On this frame, up to first order  in $1/\omega_{\text{L}}$ and due to the symmetries of the driving, the effective Hamiltonian and kick operator have the form

\begin{align}
      \label{HF_KF}   
      \hat{H}^{\text{F}}&= \hat{S}^{(0)}(\omega_0+  \hat{X})+\hat{H}_{\text{B}} \ ,\\
	\hat{K}(t)&= \hat{M}(t) \left( \omega_0 + \hat{X}   \right) \label{kick}
     \ ,
\end{align}
where
$\hat{M}(t)= \sum^{\infty}_{l\neq0}\hat{S}^{(l)}\frac{e^{\mathrm{i}l\omega_{\text{L}}t}}{\mathrm{i}\omega_{\text{L}}l}$ 
and $\hat{S}^{(l)}$ is the $l$-Fourier component of operator $\hat{S}$ in the rotating frame.

Based on the particular form of~\eqref{HF_KF},~\eqref{kick} and decomposition \eqref{Effe_Evol}, the evolution operator in the rotating frame can be seen as the product of three system-state-dependent displacement operators (polaron-type transformations) and some time-dependent phases.
With the help of the spectral decomposition of system operators $\hat{O}$ into projectors $P_{O}^{m} = \ket{O_m}\bra{O_m}$ associated to each eigenvalue $\{O_m\}$, the evolution operator $\hat{U}(t,0)=\hat{\mathcal{U}}(t)\hat{U}_{\text{R}}(t,0)$ of Eq.~\eqref{Effe_Evol} can be written in terms of the propagator in the rotating frame
\begin{align}
      \label{Evol_final}   
	 \hat{U}_{\text{R}}(t,0)=\sum_{\bm{n}} e^{-i\Omega_{\bm{n}}(t)}e^{i \text{Im}[\chi_{\bm{n}}(t)] } e^{-i H_B t} \hat{\mathcal{G}}_{\bm{n}}(t)\hat{D}[\bm{\Lambda}_{\bm{n}}(t)]
 \ .           
\end{align}
The multi-index $\bm{n}=(n_1,n_2,n_3)$
labels the eigenstates of the operators $\hat{M}(t)$, $\hat{S}^{(0)}$ and $\hat{M}(0)$, respectively.

%
\begin{figure}[htp]
  \includegraphics[width=1 \linewidth]{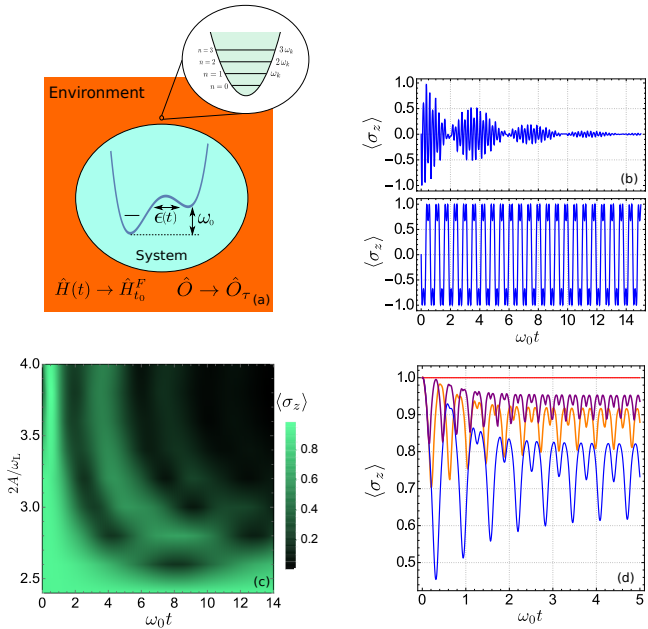}%
\caption{(a) Diagram of the spin-boson model with time dependent tunneling amplitude $\epsilon (t)=A\cos(\omega_{\text{L}}t)$, where the evolution of an observable $\hat O$ due to a driven Hamiltonian $H(t)$ can be interpreted as the evolution of a continuous family of observables $\hat{O}_{\tau}$ due to an effective static Hamiltonian $H_{t_0}^F$. (b) Expectation value of $\sigma_z$ in the laboratory frame for different amplitudes of driving strength. In the upper plot $2A/\omega_{\text{L}}\approx 3.83$ and in the lower plot $2A/\omega_{\text{L}}\approx 2.4$. Other parameters are $\omega_{\text{L}}= 10 \omega_0$, $\omega_c=0.9 \omega_0$, $\lambda = 0.15 \omega_0$ and temperature $T_{\beta}=\omega_0$. The system is initially in the state $\ket{-}_y$.
(c) Density plot of the upper envelope of expectation value $\sigma_z$ in the lab frame as a function of time and the ratio $2A/\omega_{\text{L}}$. Same parameters as (b).
(d) Expectation value of $\sigma_z$ in the rotating frame for different frequencies of the driving. Blue line corresponds to $\omega_{\text{L}}=10\omega_0$, orange to $\omega_{\text{L}}=15\omega_0$, purple to $\omega_{\text{L}}=20\omega_0$ and red to $\omega_{\text{L}} \rightarrow \infty$, where only the zeroth order term has an effect. Other parameters are  $\omega_c=0.9 \omega_0$, $\lambda = 0.5 \omega_0$, $T_{\beta}=7\omega_0$, $2A/\omega_{\text{L}}\approx 2.4$ and the system is initially in the excited state $\ket{+}_z$. 
}
\label{Fig:1}
\end{figure}
\noindent $\hat{\mathcal{G}}_{\bm{n}}(t)= P^{n_{1}}_{M(t)}P^{n_{2}}_{S^{(0)}}P^{n_{3}}_{M(0)}$ is the product of three system projectors and  the displacement operator is defined by $
\hat{D}  \left[  \bm{\mu} \right]   =e^{\sum^{N}_{k=1} \mu_k a^{\dagger}_k - \mu_k^* a_k}
$,
where $\bm{\mu}=(\mu_1,\dots,\mu_N)$. As a result of the product of the polaron-type transformations in Eq.~\eqref{Effe_Evol}, we obtain
a net displacement of the bath $\bm{\Lambda}_{\bm{n}}(t)=\bm{\alpha}_{n_{1}}(t)+ \bm{\vartheta}_{n_{2}}(t)-\bm{\alpha}_{n_{3}}(0)$ and the complex phase by $
\chi_{\bm{n}}(t)= \bm{\alpha}_{n_{1}}(t)\cdot \bm{\vartheta}_{n_{2}}^*(t)-[\bm{\alpha}_{n_{1}}(t)+\bm{\vartheta}_{n_{2}}(t)] \cdot \bm{\alpha}_{n_{3}}^*(0)$, with 
$
\alpha^{m}_{k}(t)=- i M_{m}(t) g_k e^ {i \omega_k t}$ and  $\vartheta^{n_2}_{k}(t)=\frac{S^{(0)}_{n_2}g_{k}}{\omega_{k}}(1-e^{\mathrm{i}\omega_{k}t})$. 
Additionally, we have defined the time-dependent phase
$\Omega_{\bm{n}}(t)=\omega_0\left[M_{n_1}(t)+S_{n_2}^{(0)}t-M_{n_3}(0)\right]-\eta_{n_2}(t)$, where $\eta_{n_2}(t)=\left( S_{n_2}^{(0)}\right)^{2}\sum^{N}_{k=1}\left(\frac{g_{k}}{\omega_{k}}\right)^{2}(\omega_{k}t-\sin\omega_{k}t)$. This is a general treatment and could be used in a large variety of situations beyond the scope of this letter due to the arbitrariness in the form of operators $\hat{S}$ and $\hat{V}$ in~\eqref{drivendephasingGeneral}.

In the rotating frame, the density matrix describing the dynamics of both the system and the bath has the form 
$
\hat{\rho}(t)=\hat{U}_{\text{R}}(t,0)\hat{\rho}(0)\hat{U}_{\text{R}}^{\dagger}(t,0)
$. Our approach allows us to calculate the time evolution of any initial state of the total system. For simplicity, we take $\hat{\rho}(0)=\hat{\rho}_{\text{S}}(0)\otimes \hat{\rho}_{\text{B}}(0)$, where $\hat{\rho}_{\text{B}}(0)=e^{-\beta H_{\text{B}}}/Z_{\beta}$ is a thermal state, $Z_{\beta}=\text{Tr}(e^{-\beta H_{\text{B}}})$  the partition function, and $\beta=1/T_{\beta}$ the inverse temperature. The reduced density matrix of the system obtained by tracing out the bath is
\begin{align}
      \label{Density_matrixfinal}   
	\rho_{\text{S}}(t)= \sum_{\bm{n},\bm{\tilde{n}}}e^{i\theta_{\bm{n},\bm{\tilde{n}}}(t)}e^{-\delta_{\bm{n},\bm{\tilde{n}}}(t)}\hat{\mathcal{G}}_{\bm{n}}(t)\rho_{\text{S}}(0)\hat{\mathcal{G}}^{\dagger}_{\bm{\tilde{n}}}(t)
      \ ,
\end{align}
where 
$
\theta_{\bm{n},\bm{\tilde{n}}}= \Omega_{\bm{\tilde{n}}} -\Omega_{\bm{n}}
+  \text{Im}(\chi_{\bm{n}}) 
 -  \text{Im}(\chi_{\bm{\tilde{n}}})       
 +  \text{Im}\left[  \bm{\Lambda}_{\bm{n}} \cdot \bm{\Lambda}_{\bm{\tilde{n}}}^*    \right] 
$. The dissipative effects of the bath on the system are contained on 
$
\delta_{\bm{n},\bm{\tilde{n}}}=\frac{1}{2}\sum^{N}_{k=1}|\Lambda_{k}^{\bm{n}}-\Lambda_{k}^{\bm{\tilde{n}}}|^2\coth(\beta \omega_k/2)
$. The explicit derivation of the reduced density operator of Eq.~\eqref{Density_matrixfinal} is discussed in the supplemental material~\cite{SupplementalInfo}.

\textit{Application to the spin-boson model:-}
In this section we apply our formalism to the spin boson model (Fig.~\ref{Fig:1}~a)
\begin{align}
      \label{drivendephasing}
      	\hat{H}(t)= \omega_0 \sigma_z + A\cos(\omega_{\text{L}}t)\sigma_x + \hat{H}_{\text{B}}
      	+ \sigma_z \hat{X} ,      
\end{align}
which is a paradigm of quantum dissipation~\cite{Weiss2008,Breuer2007,Thorwart2007,Grifoni2004,Hausinger2010,Thorwart2009}. Whereas we present an analytical solution, previous works have numerically explored the dissipative dynamics of
the spin boson model with a monochromatic driving on the bias term proportional to $\sigma_z$.  In this case,
signatures of coherent destruction of tunneling appear in the dynamics of the population inversion $\langle \sigma_z\rangle$ both in the Markovian \citep{Hausinger2010} as well as in the non-Markovian \citep{Thorwart2009} regimes.

From Eq.~\eqref{drivendephasingGeneral} we can identify the operators $\hat{S}=\sigma_z$ and $\hat{V}=\sigma_x$. After some algebra, we can write the operator
$\hat{M}(t) = f_t \,\sigma_z - h_t \,\sigma_y$ appearing in Eq.~\eqref{kick}, where
$
f_t=\sum^{\infty}_{m=2}\mathcal{J}_{m}(\tfrac{2A}{\omega_{\text{L}}})\frac{2\sin (m\omega_{\text{L}} t)}{m\omega_{\text{L}}}$ for even $m$, and
$
h_t=\sum^{\infty}_{m=1}\mathcal{J}_{m}(\tfrac{2A}{\omega_{\text{L}}})\frac{2\cos (m\omega_{\text{L}} t)}{m\omega_{\text{L}}}
$ for odd $m$. In these expressions, $ \mathcal{J}_{l}(x)$ is the $l$-th order Bessel function the first kind. The convergence of the results presented below was extensively tested by means of comparison with numerically exact solutions computed with the hierarchy of equations of motion \cite{Tanimura89}.

Up to first order in $1/\omega_\text{L}$, the quasienergies of the driven spin are given by the eigenvalues of the effective Hamiltonian \eqref{HF_KF} with $\hat{X}=0$. Therefore, the zeros of the Bessel function $\mathcal{J}_0(\tfrac{2A}{\omega_{\text{L}}})$ determine the occurrence of degeneracies in the Floquet spectrum as a consequence of the parity symmetry $t\mapsto t+T/2$ and $\sigma_x\mapsto -\sigma_x$ in the extended Hilbert space.
In the context of driven quantum systems, this phenomenon is known as coherent destruction of tunneling or dynamical localization~\cite{Hanggi1991,Grifoni1998}. In Fig.~\ref{Fig:1} b), the expectation value of $\sigma_z$ is shown for two specific values of $2A/\omega_{\text{L}}$ and an Ohmic spectral density $J(\omega)=  \lambda \omega e^{-\omega/\omega_{\text{c}}}$ for the bath. Although this form is used in the remainder of the letter, our solution is valid for an arbitrary spectral density.

The ratio $2A/\omega_{\text{L}}$ may be chosen to match the zeros (extrema) of Bessel function $\mathcal{J}_0(\tfrac{2A}{\omega_{\text{L}}})$, which are associated to minima (maxima) of the relaxation rate of $\sigma_z$. The value $2A/\omega_{\text{L}}\approx 3.83$ (upper panel in Fig.~\ref{Fig:1} b), corresponds to the second maximum of $\mathcal{J}_0(\tfrac{2A}{\omega_{\text{L}}})$, so that the dissipative effect of the driving, in an originally dephasing environment, can be appreciated best. Besides, collapses and revivals characteristic of driven closed systems are also apparent.     
When $2A/\omega_{\text{L}}\approx 2.4$ (lower panel in Fig.~\ref{Fig:1} b), the first zero of $\mathcal{J}_0(\tfrac{2A}{\omega_{\text{L}}})$ is matched and the effective Hamiltonian \eqref{HF_KF} only contains the free bath term. The evolution is then dominated by the kick operator and the system is effectively decoupled from the environment. Therefore, $\sigma_z$ oscillates between constant values without decaying, thus reproducing the behavior of continuous wave dynamical decoupling setups \citep{Xu2012, Hirose2012, Macquarrie2015,Monteiro2015}.  Fig.~\ref{Fig:1} c) illustrates the transition between the two limits, providing evidence of the high tunability of the effect of the driving that is available.
 
\textit{Effect of the first order term in $1/\omega_{\text{L}}$:-} Up to zeroth order in the period ($\omega_{\text{L}} \rightarrow \infty$) the effective Hamiltonian is given by \eqref{HF_KF} and the kick operator vanishes $\hat{K}(t)=0$. In such a case, the populations should stay constant in the rotating frame and the observation of any population transfer is a direct consequence of including the first order term in our treatment. This effect is shown in Fig.~\ref{Fig:1}.d), where the expectation value of $\sigma_z$ in the rotating frame is depicted for a system initially in the excited state and different values of the driving frequency $\omega_{\rm L}$. As the frequency is lowered, a cross-over between a dephasing behavior and a dissipative one can be observed. Besides the decay of populations, there is also a presence of oscillations. Parameters are chosen to match the first zero of the Bessel function ($2A/\omega_{\text{L}}\approx 2.4$), so that in this case the dissipative behavior is caused by the kick operator \eqref{kick} alone.

\begin{figure}
  \includegraphics[width=1 \linewidth]{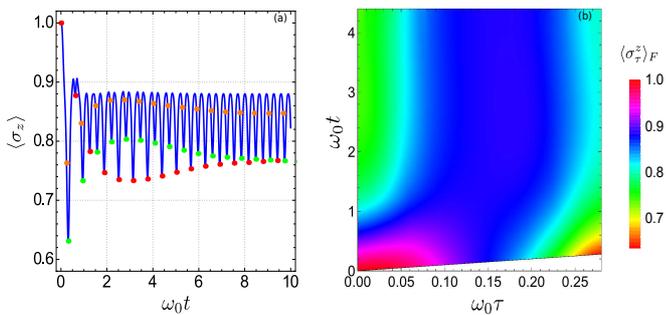}%
\caption{(a) Expectation value of $\sigma_z$ (blue solid line) for Hamiltonian \eqref{drivendephasing}. Dots show the stroboscopic simulation of static Hamiltonian \eqref{FM_Hamiltonian}: red dots correspond to $\tau=0$, green to $\tau=\pi/{\omega_{\text{L}}}$, orange to $\tau =\pi/(1.3\, \omega_{\text{L}})$. Other parameters are $\omega_{\text{L}}= 11 \omega_0$, $2 A = 2.7 \omega_{\text{L}} $, $\lambda = 0.5 \omega_0$, $\omega_c= 1.3 \omega_0$, $T_\beta=3.5\omega_0$ and $t_0=0$. The system is initially in the excited state $\ket{+}_z$. (b) Density plot of the stroboscopic simulation of $\sigma_z$ as a function of the stroboscopic parameter $\tau$ and time $t$. For each $\tau$, time can only take discrete values $\tau+nT$, and an interpolation between dots has been applied.  Parameters as in (a).}
\label{Fig:2}
\end{figure}

\textit{Stroboscopic simulation:-} For any arbitrary observable $\hat O$ and $t_1=t_0$ [see Eq.~\eqref{F_theo}], its expectation value at the stroboscopic times $t_{0n}=t_0+nT$ is simply given in terms of the Floquet Hamiltonian by $\expval{\hat{O}(t_{0n})}=\expval{\hat{O} (t_{0n})}_{F} \equiv \expval{e^{i \hat{H}_{t_0}^{\text{F}}nT}\hat{O} e^{-i \hat{H}_{t_0}^{\text{F}}nT}}$. 
This property can be extended to arbitrary stroboscopic times $\tau_n=\tau+nT$, where $t_0\leq\tau\leq t_0+T$, with help of the continuous family of observables $\hat O_{\tau} =e^{i \hat{K}_{t_0}^{\text{F}}(\tau)}\hat O e^{-i \hat{K}_{t_0}^{\text{F}}(\tau)}$.
This definition allows us to interpret the expectation value of an observable $\hat O$ at any time $\tau_n$ as the evolution of observable $\hat O_{\tau}$ under the static Hamiltonian $\hat{H}^{\text{F}}_{t_0}$, following $ \expval{\hat O(\tau_n )} = \expval{\hat O_{\tau}(\tau_n)}_{F}$. This provides us with the opportunity to simulate static Hamiltonians and observables by judiciously controlling the external driving of a simpler system. We stress that this interpretation applies in general for any periodically driven system.

For our example of the spin-boson model, the Floquet Hamiltonian \eqref{F_theo} corresponding to \eqref{drivendephasing} has the form
\begin{align}
      \label{FM_Hamiltonian}
            \hat{H}^{\text{F}}_{t_0}&= \mathcal{J}_{0}\left( \tfrac{2A}{\omega_{\text{L}}}\right)\sigma_z\left(\omega_0 + \hat{X}   \right)+(f_{t_{0}}\sigma_z+h_{t_0}\sigma_y)\dot{\hat{X}}
            \\&
            +2h_{t_0}
             \mathcal{J}_{0}\left( \tfrac{2A}{\omega_{\text{L}}}\right)\sigma_x\left(\omega_0 + \hat{X}   \right)^2+\hat{H}_{\text{B}} \nonumber
        \ ,
\end{align}
where $\dot{\hat{X}}= i \left[ \hat{H}_{\text{B}}, \hat{X}\right]= i\, \sum_{k=1}^{N} g_k \,\omega_k \left(  \hat{a}_k^{\dagger}  -  \hat{a}_k  \right)$. In addition, the stroboscopic kick operator~\eqref{F_theo} is given by 
$
\hat{K}^{\text{F}}_{t_0}(t)=\tilde{M}_{t_0}(t) \left( \omega_0 + \hat{X}   \right)
$, where~$\tilde{M}_{t_0}(t)= \widetilde{f}_{t_0}(t)\sigma_z - \widetilde{h}_{t_0}(t)\sigma_y $, $\widetilde{f}_{t_0}(t)=f_t-f_{t_0}$ and $\widetilde{h}_{t_0}(t)=h_t-h_{t_0}$. In this Hamiltonian, the parametric freedom on $A$ and $t_0$ allows us to vary the specific form of Eq.~\eqref{FM_Hamiltonian}. In particular, the strong coupling limit, i.e.,  $f_{t_{0}},h_{t_{0}} \sim \mathcal{J}_{0}\left( \tfrac{2A}{\omega_{\text{L}}}\right)$, is accessible in this case, where the usual Born-Markov master equation is not valid and polaron dynamics play an important role. This can be assessed in the evolution of the system observable $\hat{O}=\sigma_z$, whose associated continuous family is $\sigma^z_{\tau}=e^{i \hat{K}^{\text{F}}_{t_0}(\tau)}\sigma_z e^{-i \hat{K}^{\text{F}}_{t_0}(\tau)}$. This family involves such combination of system and bath operators that its expectation value, which ranges between -1 and +1, can be interpreted as a measure of polaron coherence. This can be best understood in the case $\widetilde{f}_{t_0}(\tau)\simeq0$ ($t_0=0$ and $\tau\simeq0.14/ \omega_0$), where the operator has the simple form $\sigma^z_{\tau}= \sigma_z \, e^{2 \,i\,   \widetilde{h}_{t_0}(\tau)(\omega_0 + \hat{X})\sigma_y}$. For a polaron state $\ket{+}_y\ket{\widetilde{\bm{h}}(\tau)}\pm\ket{-}_y\ket{-\widetilde{\bm{h}}(\tau)}$, where $\hat{D}[\widetilde{\bm{h}}(\tau)]\ket{\bm{0}}  \equiv \ket{\widetilde{\bm{h}}(\tau)}$ is a coherent state of the bath with $\widetilde{h}_k(\tau)=2\mathrm{i}\widetilde{h}_{t_0}(\tau)g_k$, the expectation value of $\sigma^z_{\tau}$ takes the extrema $\pm 1$. 
The function $ \widetilde{h}_{t_0}(\tau)$ is a measure of the polaron displacement from the center of the environmental phase space.

These ideas are exemplified in Fig.~\ref{Fig:2}. Fig.~\ref{Fig:2}~a) shows the dynamics of observable $\left\langle \sigma_z \right\rangle $ for the driven system \eqref{drivendephasing} in a continuous blue line. The dot series correspond to different values of the stroboscopic parameter $\tau$, corresponding to observables $\sigma^{z}_{\tau}$ under the effect of the static Hamiltonian~\eqref{FM_Hamiltonian}.  Fig.~\ref{Fig:2}~b) shows the evolution of the whole family of observables as a function of time.  Our choice of parameter values corresponds to an effective Hamiltonian with strong system-bath coupling, such that polaron dynamics plays an important role. Indeed, for an initial state $\ket{+}_z$, the system-bath state is gradually transformed into the polaron associated to the observable $\sigma^{z}_{\tau}$ for $\tau\simeq0.12 / \omega_0$. This stroboscopic simulation provides a unique insight into these dynamics under the whole range of system-bath couplings. Alternative choices of $t_0$ are associated to additional, highly non-trivial Hamiltonians, so that this procedure constitutes a flexible tool for the simulation of a large class of open quantum systems.

\textit{Conclusion and outlook:-} We have obtained the dynamics of a driven dissipative system valid to all orders in the system-bath coupling using a high-frequency expansion. The driving introduces nontrivial effects on the relaxation of populations of the system which can be accurately controlled by choice of the driving parameters. Our approach goes beyond usual studies based on weak coupling master equation and Markovian regime. Based on this solution, we also proposed a method to simulate the dynamics of non trivial static Hamiltonians for strong and weak coupling regimes. Our method can be generalized to a large class of driven-dissipative systems. Just to mention some examples, in quantum optics, the Eq.~\eqref{drivendephasingGeneral} is a driven Rabi Hamiltonian for $N=1$. Its multimode version with $N>1$ can be realized in circuit QED~\cite{Baust2016}. In cavity optomechanics, one can use $\hat{S}=\hat{b}^\dagger\hat{b}$ and $\hat{V}=\hat{b}^\dagger+\hat{b}$ to investigate non-Markovian effects on continuous variable quantum state processing~\cite{Schmidt2012}.
In the context of many-body systems, we anticipate that our method can be used as a platform to simulate systems with exotic effective interactions due to the effect of driving.

\acknowledgments{%
  V. M. B. acknowledges fruitful discussions with A. Chia and G. Engelhardt. The authors acknowledge financial support through DFG Grants No. BRA 1528/7, No. BRA 1528/8, No. BRA 1528/9, No. SFB 910, No. GRK 1558 (S. R., J. C. and T. B.), and through the Singapore Ministry of Education Academic Research Fund Tier 3 (Grant No. MOE2012-T3-1-009), National Research Foundation (NRF) Singapore and the Ministry of Education, Singapore under the Research Centres of Excellence program (V. M. B. and D. G. A.).%
 }

\bibliographystyle{phaip}
\bibliography{Mybib}

\begin{thebibliography}{10}

\bibitem{Shirley1965}
J.~H. Shirley,
\newblock Phys. Rev. {\bf 138}, B979 (1965).

\bibitem{Sambe1973}
H.~Sambe,
\newblock Phys. Rev. A {\bf 7}, 2203 (1973).

\bibitem{Grifoni1998}
M.~Grifoni and P.~H{\"a}nggi,
\newblock Phys. Rep. {\bf 304}, 229 (1998).

\bibitem{Bukov2015}
M.~Bukov, L.~D'Alessio, and A.~Polkovnikov,
\newblock Advances in Physics {\bf 64}, 139 (2015).

\bibitem{Eckardt2015}
A.~Eckardt and E.~Anisimovas,
\newblock New J. Phys. {\bf 17}, 093039 (2015).

\bibitem{Hanggi1991}
F.~Grossmann, T.~Dittrich, P.~Jung, and P.~H{\"a}nggi,
\newblock Phys. Rev. Lett. {\bf 67}, 516 (1991).

\bibitem{Hanggi1992}
F.~Grossmann and P.~H\"{a}nggi,
\newblock Europhys. Lett. {\bf 18}, 571 (1992).

\bibitem{Shevchenko20101}
S.~Shevchenko, S.~Ashhab, and F.~Nori,
\newblock Phys. Rep. {\bf 492}, 1  (2010).

\bibitem{Bastidas2014}
V.~M. Bastidas, P.~P\'erez-Fern\'andez, M.~Vogl, and T.~Brandes,
\newblock Phys. Rev. Lett. {\bf 112}, 140408 (2014).

\bibitem{Bastidas2012}
V.~M. Bastidas, C.~Emary, B.~Regler, and T.~Brandes,
\newblock Phys. Rev. Lett. {\bf 108}, 043003 (2012).

\bibitem{Hone2009}
D.~W. Hone, R.~Ketzmerick, and W.~Kohn,
\newblock Phys. Rev. E {\bf 79}, 051129 (2009).

\bibitem{HanggiFluctuation}
P.~Talkner, M.~Campisi, and P.~H\"{a}nggi,
\newblock J. Stat. Mech. {\bf 2009}, P02025 (2009).

\bibitem{Weiss2016}
M.~Carrega, P.~Solinas, M.~Sassetti, and U.~Weiss,
\newblock Phys. Rev. Lett. {\bf 116}, 240403 (2016).

\bibitem{Hong2015}
C.~Chen, J.-H. An, H.-G. Luo, C.~P. Sun, and C.~H. Oh,
\newblock Phys. Rev. A {\bf 91}, 052122 (2015).

\bibitem{Kohler1997}
S.~Kohler, T.~Dittrich, and P.~H{\"a}nggi,
\newblock Phys. Rev. E {\bf 55}, 300 (1997).

\bibitem{Szczygielski2013}
K.~Szczygielski, D.~Gelbwaser-Klimovsky, and R.~Alicki,
\newblock Phys. Rev. E {\bf 87}, 012120 (2013).

\bibitem{Szczygielski2014}
K.~Szczygielski,
\newblock J. Math. Phys. {\bf 55} (2014).

\bibitem{Hausinger2010}
J.~Hausinger and M.~Grifoni,
\newblock Phys. Rev. A {\bf 81}, 022117 (2010).

\bibitem{Thorwart2009}
J.~Eckel, J.~H. Reina, and M.~Thorwart,
\newblock New J. Phys {\bf 11}, 085001 (2009).

\bibitem{Grifoni2004}
M.~C. Goorden, M.~Thorwart, and M.~Grifoni,
\newblock Phys. Rev. Lett. {\bf 93}, 267005 (2004).

\bibitem{Breuer2007}
H.-P. Breuer and F.~Petruccione,
\newblock {\em {The Theory of Open Quantum Systems}},
\newblock Oxford University Press, Oxford, 2007.

\bibitem{Weiss2008}
U.~Weiss,
\newblock {\em {Quantum dissipative systems}}, volume~10,
\newblock World Scientific, 1999.

\bibitem{Shirai2015}
T.~Shirai, T.~Mori, and S.~Miyashita,
\newblock Phys. Rev. E {\bf 91}, 030101 (2015).

\bibitem{Liu2015}
D.~E. Liu,
\newblock Phys. Rev. B {\bf 91}, 144301 (2015).

\bibitem{Cerrillo2014}
J.~Cerrillo and J.~Cao,
\newblock Phys. Rev. Lett. {\bf 112}, 110401 (2014).

\bibitem{Brandes2005}
T.~Vorrath and T.~Brandes,
\newblock Phys. Rev. Lett. {\bf 95}, 070402 (2005).

\bibitem{Brandes2004}
R.~Aguado and T.~Brandes,
\newblock Phys. Rev. Lett. {\bf 92}, 206601 (2004).

\bibitem{Schindler2013}
P.~Schindler et~al.,
\newblock Nat. Phys. {\bf 9}, 361 (2013).

\bibitem{Mintert2015b}
F.~Haddadfarshi, J.~Cui, and F.~Mintert,
\newblock Phys. Rev. Lett. {\bf 114}, 130402 (2015).

\bibitem{Thorwart2007}
F.~Nesi, E.~Paladino, M.~Thorwart, and M.~Grifoni,
\newblock Phys. Rev. B {\bf 76}, 155323 (2007).

\bibitem{Goldman2014}
N.~Goldman and J.~Dalibard,
\newblock Phys. Rev. X {\bf 4}, 031027 (2014).

\bibitem{Rahav2003}
S.~Rahav, I.~Gilary, and S.~Fishman,
\newblock Phys. Rev. A {\bf 68}, 013820 (2003).

\bibitem{Itin2015}
A.~P. Itin and M.~I. Katsnelson,
\newblock Phys. Rev. Lett. {\bf 115}, 075301 (2015).

\bibitem{Mikami2015}
T.~Mikami et~al.,
\newblock Phys. Rev. B {\bf 93}, 144307 (2016).

\bibitem{Mananga2011}
E.~S. Mananga and T.~Charpentier,
\newblock J. Chem. Phys. {\bf 135} (2011).

\bibitem{Kuwahara201696}
T.~Kuwahara, T.~Mori, and K.~Saito,
\newblock Ann. Phys. {\bf 367}, 96  (2016).

\bibitem{Lopez2013}
A.~L{\'o}pez, A.~Scholz, Z.~Z. Sun, and J.~Schliemann,
\newblock Eur. Phys. J.B {\bf 86}, 1 (2013).

\bibitem{SupplementalInfo}
{See supplemental material for additional details.}

\bibitem{Tanimura89}
Y.~Tanimura and R.~Kubo,
\newblock Journal of the Physical Society of Japan {\bf 58}, 101 (1989).

\bibitem{Xu2012}
X.~Xu et~al.,
\newblock Phys. Rev. Lett. {\bf 109}, 070502 (2012).

\bibitem{Hirose2012}
M.~Hirose, C.~D. Aiello, and P.~Cappellaro,
\newblock Phys. Rev. A {\bf 86}, 062320 (2012).

\bibitem{Macquarrie2015}
E.~R. MacQuarrie, T.~A. Gosavi, S.~A. Bhave, and G.~D. Fuchs,
\newblock Phys. Rev. B {\bf 92}, 224419 (2015).

\bibitem{Monteiro2015}
J.~E. Lang, R.~B. Liu, and T.~S. Monteiro,
\newblock Phys. Rev. X {\bf 5}, 041016 (2015).

\bibitem{Baust2016}
A.~Baust et~al.,
\newblock Phys. Rev. B {\bf 93}, 214501 (2016).

\bibitem{Schmidt2012}
M.~Schmidt, M.~Ludwig, and F.~Marquardt,
\newblock New Journal of Physics {\bf 14}, 125005 (2012).

\end{thebibliography}


 \begin{widetext}
 

\section*{\large Supplemental Material }

\subsection{High Frequency Expansions}\label{Sup:HFE}

We present the the Floquet-Magnus expansion and the (van Vleck) high-frequency expansion (HFE). For a more detailed description, refer to \citep{Goldman2014,Bukov2015,Eckardt2015,Mananga2011,Rahav2003,Itin2015,Mikami2015}.
For a time-dependent periodic Hamiltonian $\hat{H}(t)=\hat{H}(t+T)$ with period $T=2 \pi/\omega_{\text{L}}$ one can express the Hamiltonian in terms of its Fourier series $\hat{H}(t)= \sum_{n= -\infty}^{\infty} \hat{H}_n \,e^{\ii n \omega_{\text{L}} t}$, with $n$ a positive integer. The dynamics of such a Hamiltonian are described by the evolution operator
\begin{eqnarray}\label{evolution}
\hat{U}(t,t_0) &=& \hat{T_t} \left\lbrace  \exp \left[ - \ii \int^{t}_{t_0} d t^\prime \hat{H}(t^\prime)   \right]\right\rbrace  =  e^{-\ii \hat{K}^{\text{F}}_{t_0}(t)}e^{-\ii \hat{H}_{t_0}^{\text{F}}(t-t_0)} 
\end{eqnarray}
where $\hat{T_t}$ is the time-ordering operator and the last term is the statement of the Floquet theorem. The first few terms of the Floquet-Magnus expansion read
\begin{align}
\hat{H}_{t_0}^{\text{F}(0)} &= \dfrac{1}{T} \int^{T+t_0}_{t_0} dt \hat{H}(t)  = \hat{H}_0\\
\hat{H}_{t_0}^{\text{F}(1)}&= \dfrac{-\ii}{2 T} \int^{T+t_0}_{t_0} dt_1 \int^{t_1}_{t_0} dt_2  \left[ \hat{H}(t_1),\hat{H}(t_2)  \right] = \dfrac{1}{\omega_{\text{L}}}\sum_{n= 1}^{\infty} \dfrac{1}{n}  \left(  \left[\hat{H}_n,\hat{H}_{-n}  \right] - e^{\ii n\omega_{\text{L}} t_0} \left[\hat{H}_{n},\hat{H}_0  \right] -e^{-\ii n\omega_{\text{L}} t_0} \left[\hat{H}_{-n},\hat{H}_0  \right]  \right)  \\
\hat{K}^{\text{F}(0)}_{t_0}(t) &= 0, \\
\hat{K}^{\text{F}(1)}_{t_0}(t)&= \int^{t}_{t_0} dt'  \left(  \hat{H}(t')-\hat{H}_{\text{F}}^{(1)} \right)= \dfrac{1}{\ii \omega_{\text{L}}} \sum_{n\neq 0} \hat{H}_n \dfrac{e^{\ii n \omega_{\text{L}} t}-e^{\ii n \omega_{\text{L}} t_0}}{n}.
\end{align}
These terms constitute the first order terms of the well known Floquet-Magnus expansion, where the Floquet Hamiltonian provides a stroboscopic description of the system dynamics. Both the kick operator and the Floquet Hamiltonian carry a dependence on $t_0$.

\noindent The dependence of the expansion on $t_0$ can be removed by applying a unitary transformation that leads to the Van-Vleck expansion. Under this transformation, the evolution operator has the form $\hat{U}(t,t_0) =e^{-\ii \hat{K}(t)} e^{-\ii \hat{H}^{\text{F}} t} e^{\ii \hat{K}(t_0)}$ and the first terms of the expansion are given by
\begin{align}
\hat{H}^{\text{F}(0)} &= \dfrac{1}{T} \int^{T}_{0} dt \hat{H}(t)  = \hat{H}_0. \qquad \qquad \hat{K}^{(0)}(t) = 0, \\
\hat{H}^{\text{F}(1)}&=  \dfrac{1}{\omega_{\text{L}}}\sum_{n= 1}^{\infty} \dfrac{1}{n}   \left[\hat{H}_n,\hat{H}_{-n}  \right] . \qquad\qquad \hat{K}^{(1)}(t)= \dfrac{1}{\ii \omega_{\text{L}}} \sum_{n\neq 0} \hat{H}_n \dfrac{e^{\ii n \omega_{\text{L}} t}}{n}.
\end{align}

\subsection{Derivation of the general evolution Operator}
Let us consider the system-bath Hamiltonian
\begin{align}
      \label{drivendephasingSup}
      	\hat{H}(t)= \omega_0\hat{S} +A\cos(\omega_{\text{L}}t)\hat{V}+ \hat{H}_{\text{B}}
      	+  \hat{S} \, \hat{X} ,      
\end{align}
where
$\hat{H}_{\text{B}}=\sum_{k=1}^{N}  \omega_k \hat{a}_k^{\dagger} \hat{a}_k$  is the Hamiltonian of the bath with $N$ modes, and $\hat{X}=\sum_{k=1}^{N} g_k \left(\hat{a}_k^{\dagger} +\hat{a}_k \right)$. In a rotating frame defined by operator 
$
\hat{\mathcal{U}}(t)= e^{  -\ii \frac{A}{\,\omega_{\text{L}}} \sin(\omega_{\text{L}} t)\, \hat{V}},
$
the transformed Hamiltonian has the form $\hat{H}^{\text{R}}(t)= \hat{S}(t)(\omega_0+ \hat{X})+  \hat{H}_{\text{B}}
 $,
with $\hat{S}(t)=\hat{\mathcal{U}}^{\dagger}(t)\hat{S}\hat{\mathcal{U}}(t)=\sum_{l=-\infty}^{\infty}\hat{S}^{(l)} e^{\ii l\omega_{\text{L}}t}$, due to the periodicity of the transformation. Now we can expand the Hamiltonian
$\hat{H}^{\text{R}}(t)=\hat{H}^{\text{R}}_{0}+\sum_{l=1}^{\infty}\left(\hat{H}^{\text{R}}_{l}e^{\ii l \omega_{\text{L}}t}+\hat{H}^{\text{R}}_{-l}e^{ - \ii l \omega_{\text{L}}t}\right)$, where $\hat{H}^{\text{R}}_{0}=\hat{S}^{(0)}(\omega_0+\hat{X})+  \hat{H}_{\text{B}}$ and $\hat{H}^{\text{R}}_{l}=\hat{S}^{(l)}(\omega_0+\hat{X})$.

Since the system satisfies the algebraic condition $\hat{H}^{\text{R}}_{-l}=(-1)^l\hat{H}^{\text{R}}_{l}$, the effective Hamiltonian and kick operator read
\begin{align}
      \label{SuppHF_KF}   
      \hat{H}^{\text{F}}&= \hat{S}^{(0)}(\omega_0+  \hat{X})+\hat{H}_{\text{B}} \\
	\hat{K}(t)&= \hat{M}(t) \left( \omega_0 + \hat{X}   \right) 
,
\end{align}
where we have defined
$\hat{M}(t)= \sum^{\infty}_{l=1}\frac{\hat{S}^{(2l)}}{l\omega_{\text{L}}}\sin (2l\omega_{\text{L}} t)+ \sum^{\infty}_{l=0}\frac{2\hat{S}^{(2l+1)}}{\mathrm{i}(2l+1)\omega_{\text{L}}}\cos [(2l+1)\omega_{\text{L}} t]$.

The operators $\hat{M}(t)$, $ \hat{S}^{(0)}$ and $\hat{M}(t)$ have eigenvalues $M_{n_1}(t)$, $S^{(0)}_{n_2}$, and $M_{n_3}(0)$, respectively.
The evolution operator in the rotating frame can then be explicitly written as
\begin{align}
      \label{Evolop2}   
	 \hat{U}_{\text{R}}(t,0)&=e^{-\ii \hat{M}(t) (\omega_0 + \hat{X})} e^{- \ii \left[\hat{S}^{(0)}(\omega_0+\hat{X})+\hat{H}_{\text{B}}   \right]t} e^{\ii \hat{M}(0) (\omega_0 + \hat{X})}.
\end{align}
We can use the decomposition $ e^{-\ii H^{\text{F}} t}=\sum_{n_2} e^{-\mathrm{i}(\omega_{0}S^{(0)}_{n_2}+H_{\text{B}})t}e^{\mathrm{i}\eta_{n_2}(t)}\ket{S^{(0)}_{n_2}}\bra{S^{(0)}_{n_2}}\hat{D}[\bm{\vartheta}_{n_2}(t)]$, $\vartheta^{n_2}_{k}(t)=\frac{S^{(0)}_{n_2}g_{k}}{\omega_{k}}(1-e^{\mathrm{i}\omega_{k}t})$ and  $\eta_{n_2}(t)=\left( S_{n_2}^{(0)}\right)^{2}\sum^{N}_{k=1}\left(\frac{g_{k}}{\omega_{k}}\right)^{2}[\omega_{k}t-\sin\omega_{k}t]$, where $\hat{S}^{(0)}\ket{S^{(0)}_{n_2}}=S^{(0)}_{n_2}\ket{S^{(0)}_{n_2}}$ and we have considered the displacement operator $
D(\bm{\mu})=e^{\sum_{k=1}^{N} \mu_k a^{\dagger}_k - \mu_k^* a_k}
$,
with $\bm{\mu}=(\mu_1,\ldots,\mu_N)$. To calculate the propagator, we also need to calculate the exponential function of the kick operator $K(t)$. This can be done by using the instantaneous eigenbasis $\{\ket{M_{n_2}(t)}\}$ of the operator $\hat{M}(t)$. After some algebra, we obtain $e^{-\ii \hat{M}(t) (\omega_0 + \hat{X})}=\sum_{n_1} e^{-\mathrm{i}\omega_0 M_{n_1}(t)}\ket{M_{n_1}(t)}\bra{M_{n_1}(t)}\hat{D}[\bm{\alpha}_{n_1}(t)]$, where $\alpha^{m}_{k}(t)=- \ii M_{m}(t) g_k e^ {i \omega_k t}$.

We now define the operator $\hat{\mathcal{G}}_{\bm{n}}(t)= P^{n_{1}}_{M(t)}P^{n_{2}}_{S^{(0)}}P^{n_{3}}_{M(0)}$, where
$P_{O}^{m} = \ket{O_m}\bra{O_m}$ are projectors into the eigenstates of a given operator $\hat{O}$.  By using this, and  the identity for the product of three displacement operators $D(\bm{\alpha})D(\bm{\beta})D(\bm{\gamma})=e^{\mathrm{i}\ \text{Im}(\bm{\alpha}\cdot\bm{\beta}^{*}+(\bm{\alpha}+\bm{\beta})\cdot\bm{\gamma}^{*})}D(\bm{\alpha}+\bm{\beta}+\bm{\gamma})$, we obtain the final form of the propagator
\begin{align}
      \label{Evol_final2}         
	 \hat{U}_{\text{R}}(t,0)=\sum_{\bm{n}} e^{-i\Omega_{\bm{n}}(t)}e^{i \text{Im}[\chi_{\bm{n}}(t)] } e^{-i H_B t} \hat{\mathcal{G}}_{\bm{n}}(t)\hat{D}[\bm{\Lambda}_{\bm{n}}(t)],      
\end{align}
\noindent where $\Omega_{\bm{n}}(t)=\omega_0\left[M_{n_1}(t)+S_{n_2}^{(0)}t-M_{n_3}(0)\right]-\eta_{n_2}(t)$, $\bm{\Lambda}_{\bm{n}}(t)=\bm{\alpha}_{n_{1}}(t)+ \bm{\vartheta}_{n_{2}}(t)-\bm{\alpha}_{n_{3}}(0)$ and $
\chi_{\bm{n}}(t)= \bm{\alpha}_{n_{1}}(t)\cdot \bm{\vartheta}_{n_{2}}^*(t)-[\bm{\alpha}_{n_{1}}(t)+\bm{\vartheta}_{n_{2}}(t)] \cdot \bm{\alpha}_{n_{3}}^*(0)$.

\subsection{Calculation of the reduced density matrix of the system in the rotating frame}
The density matrix describing the dynamics of both, the system and the bath, has the form 
$
\hat{\rho}(t)=\hat{U}_{\text{R}}(t,0)\hat{\rho}(0)\hat{U}_{\text{R}}^{\dagger}(t,0).
$
Taking the trace over the bath, the reduced density matrix of the system reads:
\begin{equation}
      \label{Density_matrix2}   
	\rho_{\text{S}}(t)=\sum_{\bm{n},\bm{\tilde{n}}}e^{\ii\theta_{\bm{n},\bm{\tilde{n}}}(t)} \hat{\mathcal{G}}_{\bm{n}}(t)\rho(0)\hat{\mathcal{G}}^{\dagger}_{\bm{\tilde{n}}}(t)\left\langle D(\bm{\Lambda}_{\bm{n}}-\bm{\Lambda}_{\bm{\tilde{n}}})  \right\rangle  \,
      , 
\end{equation}
where
$
\left\langle D(\bm{\mu})  \right\rangle = \text{Tr}_{\text{B}} \lbrace D(\bm{\mu}) \rho_{\text{B}}  \rbrace  = e^{- \sum_{k=1}^{N} \frac{ \vert \mu_k \vert^2}{2}\coth(\frac{\omega_k \beta}{2}) }
$
is the expectation value of the displacement operator assuming an initial thermal state with inverse temperature $\beta$. 
The dynamical phase is given by 
$
\theta_{\bm{n},\bm{\tilde{n}}}(t)= \Omega_{\bm{\tilde{n}}}(t) -\Omega_{\bm{n}}(t)
+  \text{Im}(\chi_{\bm{n}}) 
 -  \text{Im}(\chi_{\bm{\tilde{n}}})       
 +  \text{Im}\left[  \Lambda_{\bm{n}} \cdot \Lambda_{\bm{\tilde{n}}}^*    \right] 
$. The reduced density matrix of the system has its final form 
\begin{equation}
      \label{Density_matrixfinal2}   
	\rho_{\text{S}}(t)= \sum_{\bm{n},\bm{\tilde{n}}}e^{\ii\theta_{\bm{n},\bm{\tilde{n}}}(t)}e^{-\delta_{\bm{n},\bm{\tilde{n}}}(t)}\hat{\mathcal{G}}_{\bm{n}}(t)\rho(0)\hat{\mathcal{G}}^{\dagger}_{\bm{\tilde{n}}}(t)
      .
\end{equation}

\subsubsection{Calculation of the reduced density matrix of the qubit for the spin-boson model}
In the particular case of the spin-boson model the Hamiltonian is the following $\hat{H}= \omega_0 \sigma_z + A\cos(\omega_{\text{L}}t)\sigma_x + \hat{H}_{\text{B}}
      	+ \sigma_z \hat{X}$. Operator $\hat{M}(t)$ can be easily identified as $\hat{M}(t) = f_t \,\sigma_z - h_t \,\sigma_y$, where
$
f_t=\sum^{\infty}_{m=2}\mathcal{J}_{m}(\tfrac{2A}{\omega_{\text{L}}})\frac{2\sin (m\omega_{\text{L}} t)}{m\omega_{\text{L}}}$ for even $m$, and
$
h_t=\sum^{\infty}_{m=1}\mathcal{J}_{m}(\tfrac{2A}{\omega_{\text{L}}})\frac{2\cos (m\omega_{\text{L}} t)}{m\omega_{\text{L}}}
$ for odd $m$. In these expressions, $ \mathcal{J}_{l}(x)$ is the $l$-th order Bessel function the first kind. An integral expression for the time-dependent function in $\hat{M}(t)$ can also be found. They have the form $
f_t=\int_0^t du\, \cos  \left[  \tfrac{2A}{\omega_{\text{L}}} \sin(\omega_{\text{L}}  u)    \right] -   \mathcal{J}_{0}\left(\tfrac{2A}{\omega_{\text{L}}}\right)\, t$ and  
$
h_t=-\int_{\tfrac{\pi}{2 \omega_{\text{L}}}}^t du\, \sin  \left[  \tfrac{2A}{\omega_{\text{L}}} \sin(\omega_{\text{L}}  u)    \right]
$.   
Taking the continuum limit and using the spectral density of the bath 
$
J(\omega)= \sum_k \vert g_k \vert^2 \delta(\omega-\omega_k)
$, the exponential terms can be written as
\begin{align}
      \label{phases2}   
      \delta_{\bm{n},\bm{\tilde{n}}}&=  2 \, \mathcal{J}_{0}^2(\tfrac{2A}{\omega_{\text{L}}}) (-1 + \tilde{n}_2 n_2)   \int d\omega \, J(\omega) \frac{1-\cos(\omega t)}{\omega^2} \coth(\tfrac{\omega \beta}{2}) \\ 
       & +\left[ (1-\tilde{n}_1 n_1) \eta_t^2 + (1-\tilde{n}_3 n_3) \eta_0^2      \right] \int d\omega \, J(\omega) \coth(\tfrac{\omega \beta}{2})   
     -\Delta n_1 \Delta n_3 \, \eta_t \eta_0  \int d\omega \, J(\omega)\cos(\omega t)\coth(\tfrac{\omega \beta}{2}) \nonumber  \\	
&+   \mathcal{J}_{0}(\tfrac{2A}{\omega_{\text{L}}}) \Delta n_2 \left[ \Delta n_1 \eta_t - \Delta n_2 \eta_0    \right]    
 \int d\omega \, J(\omega) \tfrac{\sin(\omega t)}{\omega}\coth(\tfrac{\omega \beta}{2})  \nonumber  \\	
\theta_{\bm{n},\bm{\tilde{n}}}&= \Omega_{\bm{\tilde{n}}}(t) -\Omega_{\bm{n}}(t) + \mathcal{J}_{0}^2(\tfrac{2A}{\omega_{\text{L}}})  \left[  (n_3+\tilde{n}_3)\Delta n_2 \,  \eta_0 - (n_2+\tilde{n}_2) \Delta n_1 \,\eta_t      \right]  \,\int d\omega \, J(\omega) \frac{1-\cos(\omega t)}{\omega}\\ 
&+ (n_3+\tilde{n}_3) \Delta n_1 \, \eta_t \eta_0 \int d\omega \, J(\omega) \sin(\omega t), \nonumber
\end{align}
where we have defined $\eta_t = \sqrt{f_t^2 + h_t^2}$ and $\Delta n_l=\tilde{n}_l-n_l$. In this particular case, the index $n_j=\pm1$ for $j\in\{1,2,3\}$.

\end{widetext}

\end{document}